\documentclass[a4paper,11pt]{article}
\usepackage{pos}
\usepackage{braket}
\usepackage{tikz}
\usepackage{hyperref}
\usepackage{bbold}

\title{QCD mesonic screening masses and restoration
        of chiral symmetry at high T}
\author*[a,b]{Davide Laudicina}
\author[c]{Mattia Dalla Brida}
\author[a,b]{Leonardo Giusti}
\author[d]{Tim Harris}
\author[b]{Michele Pepe}

\affiliation[a]{Dipartimento di Fisica, Università di Milano-Bicocca,\\
Piazza della Scienza 3, I-20126, Milano, Italy}

\affiliation[b]{INFN, sezione di Milano-Bicocca,\\
Piazza della Scienza 3, I-20126, Milano, Italy}

\affiliation[c]{Theoretical Physics Department, CERN,\\
CH-1211 Geneva 23, Switzerland}

\affiliation[d]{School of Physics and Astronomy, University of Edinburgh,\\
Edinburgh EH9 3JZ, UK}

\emailAdd{davide.laudicina@mib.infn.it}
\emailAdd{mattia.dalla.brida@cern.ch}
\emailAdd{leonardo.giusti@mib.infn.it}
\emailAdd{tharris@ed.ac.uk}
\emailAdd{michele.pepe@mib.infn.it}

\abstract{We present a strategy to study QCD non-perturbatively on the lattice at very high temperatures. This strategy exploits a non-perturbative, finite-volume, definition of the strong coupling constant to renormalize the theory. As a first application we compute the flavor non-singlet mesonic screening masses in a wide range of temperature, from $T\sim 1 $ GeV up to $\sim 160 $ GeV with three flavors in the chiral limit of QCD. Our results show very interesting features of the screening spectrum at very high temperatures. On one hand the mass splitting between the vector and the pseudoscalar screening masses is clearly visible up to the electroweak scale and cannot be explained by the known 1-loop perturbative result. On the other hand the restoration of chiral symmetry manifests itself through the degeneracy of the pseudoscalar and the scalar channels and of the vector and the axial-vector ones. This degeneracy pattern is the one expected by Ward identities associated to the presence of chiral symmetry.
\begin{flushright}
CERN-TH-2022-201
\end{flushright}
}
\FullConference{%
  The 39th International Symposium on Lattice Field Theory (Lattice2022),\\
  8-13 August, 2022 \\
  Bonn, Germany 
}

\begin{document}
\maketitle

\section{Introduction}
Quantum Chromodynamics (QCD) at finite temperature plays a fundamental r\^ole in many fields of research, from the cosmological evolution of the early universe to the interpretation of the experimental results at relativistic heavy-ion colliders in nuclear and particle physics. As the temperature increases, due to asymptotic freedom, the theory undergoes a crossover from a confined phase with hadronic degrees of freedom, in which chiral symmetry is spontaneously broken, to a deconfined and chirally symmetric phase in which the relevant degrees of freedom are quarks and gluons.

The purpose of this talk is to present the results on the mesonic screening spectrum, obtained in Ref. \cite{DallaBrida:2021ddx}, in the extremely high temperature regime, i.e. from $T\sim 1$ GeV up to $\sim 160$ GeV. With respect to that work, here we additionally report the study of the degeneracy between various channels of the screening masses in presence of chiral symmetry restoration, as suggested by the corresponding Ward Identities which are reported in Sec. \ref{sec:WI}, for recent reviews on the subject see Ref. \cite{Aarts:2022esd,Glozman:2022lda}. The exploration of such a high temperature regime has been possible thanks to the strategy implemented to renormalize the theory on the lattice. This strategy exploits the knowledge of a non-perturbative definition of the renormalized coupling defined in a finite volume. This strategy was first used in the SU(3) Yang-Mills theory where it allowed a precise determination of the Equation of State over two orders of magnitude in the temperature \cite{Giusti:2014ila,Giusti:2016iqr}.
\section{Strategy and lattice set-up}
Typically, at zero temperature the scale on the lattice is set by using a hadronic scale $M_{\textrm{had}}$. This scale is chosen to satisfy the relation $a\ll 1/M_{\textrm{had}} \ll L$ where $a$ and $L$ are the lattice spacing and the lattice extent respectively. When we simulate the theory at finite temperature the additional scale $T$ as to be accommodated on the lattice. If the temperature is much larger than the hadronic scale, the relation to be satisfied becomes
\begin{equation}
    a\ll \frac{1}{T}\ll \frac{1}{M_{\textrm{had}}}\ll L \, ,
\end{equation}
and the computational effort for such numerical simulation is prohibitively expensive.

To overcome this problem we consider a renormalisation of the theory based on a non-perturbative definition of the renormalized coupling in a finite volume, $\Bar{g}_{SF}^2(\mu_{SF})$, where $\mu_{SF}=1/L_{SF}$ and $L_{SF}$ is the lattice extent. Here, we refer to the Schr\"odinger Functional (SF) definition of the renormalized coupling \cite{Luscher:1993gh}, but other choices are possible as well. The key idea is then to relate finite volume setup with Schr\"odinger Functional boundary conditions \cite{DallaBrida:2016kgh,DallaBrida:2016uha,DallaBrida:2018rfy} with finite temperature ones with periodic boundary conditions by requiring
\begin{align}
    T \, = \, \mu_{SF} \qquad \longrightarrow \qquad L_0 \, = \, L_{SF}\,,
\end{align}
where $L_0$ is the lattice extent in the compact direction. Finally the lines of constant physics are set by fixing the value of the renormalized coupling at finite lattice spacing to be 
\begin{align}
    \Bar{g}^2_{SF}(g_0^2,a/L_0) \, = \, \Bar{g}^2_{SF}(1/L_0)\, , \qquad a/L_0 \,=\, aT\ll 1 \, .
\end{align}
The combination of these definitions with step-scaling techniques allows us to explore a wide range of temperatures without the need of simulating very large physical volumes. Moreover, the use of this strategy is supported by the fact that finite volume effects are exponentially suppressed for sufficiently large $LT$. For this reason, in our study we have always kept $LT$ between 20 and 50, see appendix C of Ref. \cite{DallaBrida:2021ddx} for the details.

In the present study, this strategy has been implemented to simulate QCD with $N_f=3$ quarks in the chiral limit at 12 values of the temperature, between $\sim 1$ GeV and 160 GeV. Monte Carlo simulations were performed over lattices with extent $L/a=288$ in the spatial directions and with 3 or 4 different values of the lattice spacings ($L_0/a=4,6,8$ and $10$) to allow a continuum limit extrapolation. We considered shifted boundary conditions in the temporal extent with shift vector $\boldsymbol{\xi}=(1,0,0)$ \cite{Giusti:2010bb,Giusti:2011kt,Giusti:2012yj,DallaBrida:2020gux}. Even if the use of shifted boundary conditions is not crucial for this work, it gives us milder discretization errors and allowed us to share the cost of generating gauge field configurations with the project which aims at the computation of the Equation of State \cite{DallaBrida:2020gux}.

At finite temperature the topological susceptibility is expected to be proportional to $T^{-8}$ in the theory with $N_f=3$ 
quarks \cite{Bonati:2015vqz,Athenodorou:2022aay,Borsanyi:2016ksw}. As a consequence, only the zero-topology sector is relevant if the temperature is sufficiently large. For this reason, since, even at the lowest temperature we simulated, the probability to visit a gauge field configuration with non-zero topology is extremely small, we restrict our calculation to the zero-topology sector.
\section{Numerical results}
The definition of mesonic screening masses is related to the large-distance behaviour of screening correlation functions of fermionic bilinears. In particular, when projecting on the lowest Matsubara frequency, the correlation functions read
\begin{align}
    C_{\mathcal{O}}(x_3) \, = \, \int dx_0 dx_1 dx_2 \braket{\mathcal{O}^a(x)\mathcal{O}^a(0)} \underset{x_3\rightarrow \infty}{\sim}  e^{-m_{\mathcal{O}}x_3} \, ,
\end{align}
where $m_{\mathcal{O}}$ is the screening mass related to the interpolating operator $\mathcal{O}^a(x)=\overline{\psi}(x)\Gamma T^a \psi (x)$ and $T^a$ are the traceless generators of the flavor group (in this work $\Gamma\,=\,\{\mathbb{1},\gamma_5,\gamma_\mu,\gamma_\mu\gamma_5\}$). The spectrum of the mesonic screening masses presents two distinct features: on one hand, for asymptotically high temperatures, all these masses are expected to approach the value $m_{\textrm{free}} = 2\pi T$, which corresponds to the energy of two free quarks, with thermal mass $\pi T$. On the other hand, if in the high temperature regime chiral symmetry is restored, a degenerate pattern arises leading to chiral multiplets. The 1-loop order correction to the free theory value has been computed by matching the dimensionally reduced effective field theory to QCD at 1-loop in perturbation theory. It is independent of the mesonic operator $\mathcal{O}^a$ and its expression reads \cite{Laine:2003bd}
\begin{align}
\label{eq:pt}
    m_{\mathcal{O}}^{PT} \, = \, 2\pi T \left( 1+0.032739961 \cdot g^2\right)\, .
\end{align}
For this reason, besides the theoretical and physical interest, related to the fact that these masses can be used as ideal probes of chiral symmetry restoration in the quark-gluon plasma, our lattice calculation represents a test of the reliability of the 1-loop order perturbative result over more than two orders of magnitude in the temperature.
\subsection{Vector-Pseudoscalar spectrum}
In this section we discuss the main result that we obtained for the vector and the pseudoscalar spectrum. The difference between these two masses encodes spin-dependent terms of the screening masses. The effective field theory analysis predicts spin-dependent terms to be $O(g^4)$ in the renormalized coupling \cite{Koch:1992nx,Hansson:1991kb}.

In order to compare with the perturbative result, we parametrize our findings by using the perturbative definition of the renormalized coupling, in the $\overline{\textrm{MS}}$ scheme, evaluated at the renormalization scale $\mu=2\pi T$
\begin{align}
    \frac{1}{\hat{g}^2(T)}\, = \, \frac{9}{8\pi^2} \ln{\frac{2\pi T}{\Lambda_{\overline{\textrm{MS}}}}} + \frac{4}{9\pi^2}\ln{\left( 2\ln{\frac{2\pi T}{\Lambda_{\overline{\textrm{MS}}}}} \right)} \, ,
\end{align}
where $\Lambda_{\overline{\textrm{MS}}}$ is taken from Ref. \cite{Bruno:2017gxd}. Notice that the renormalized coupling is just a convenient function of the temperature, used to parametrize our results, driven by the observed leading logarithmic dependence on $T$ of the screening masses in the effective field theory.

\begin{figure}
    \centering
    \includegraphics[width=0.4\textwidth]{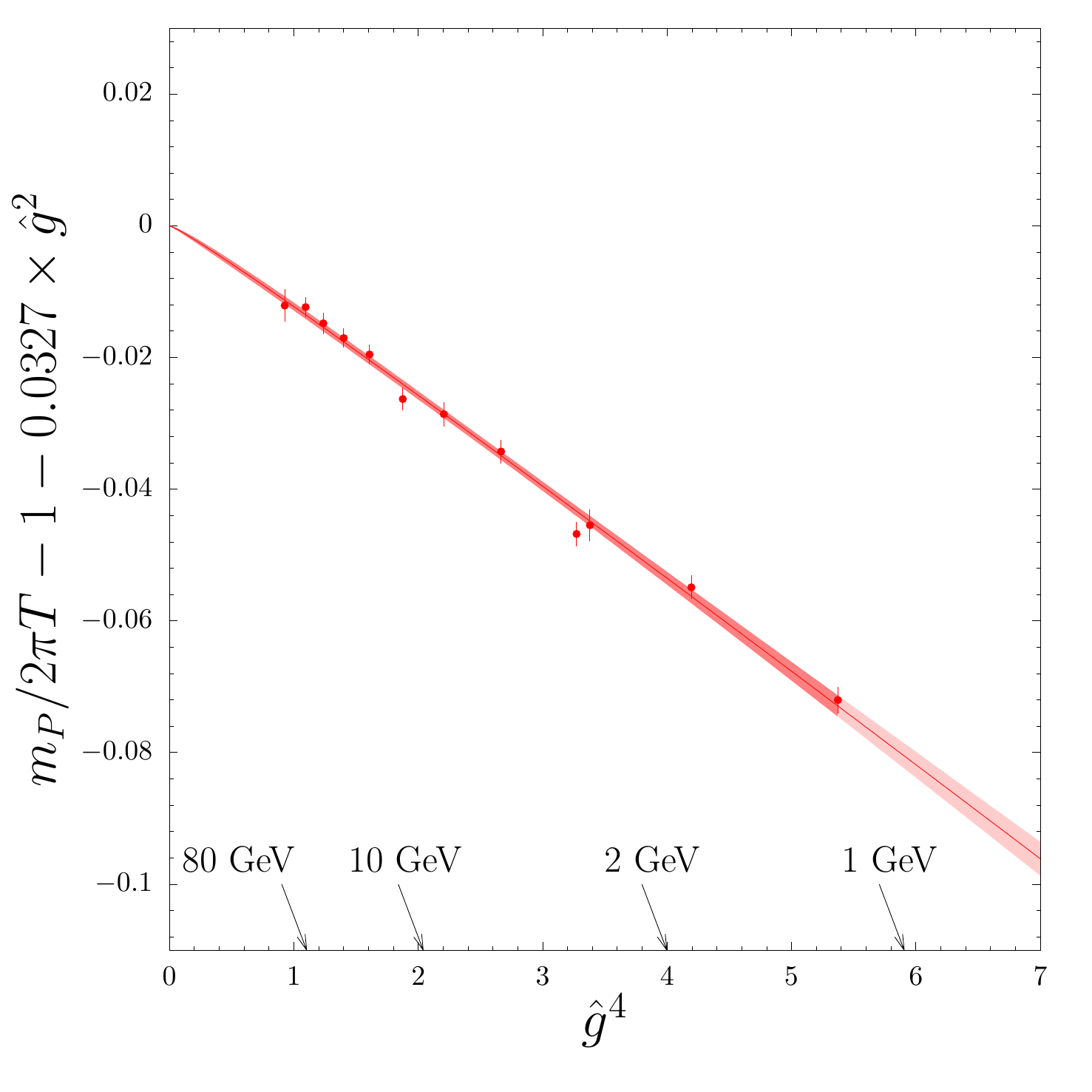}
    \includegraphics[width=0.4\textwidth]{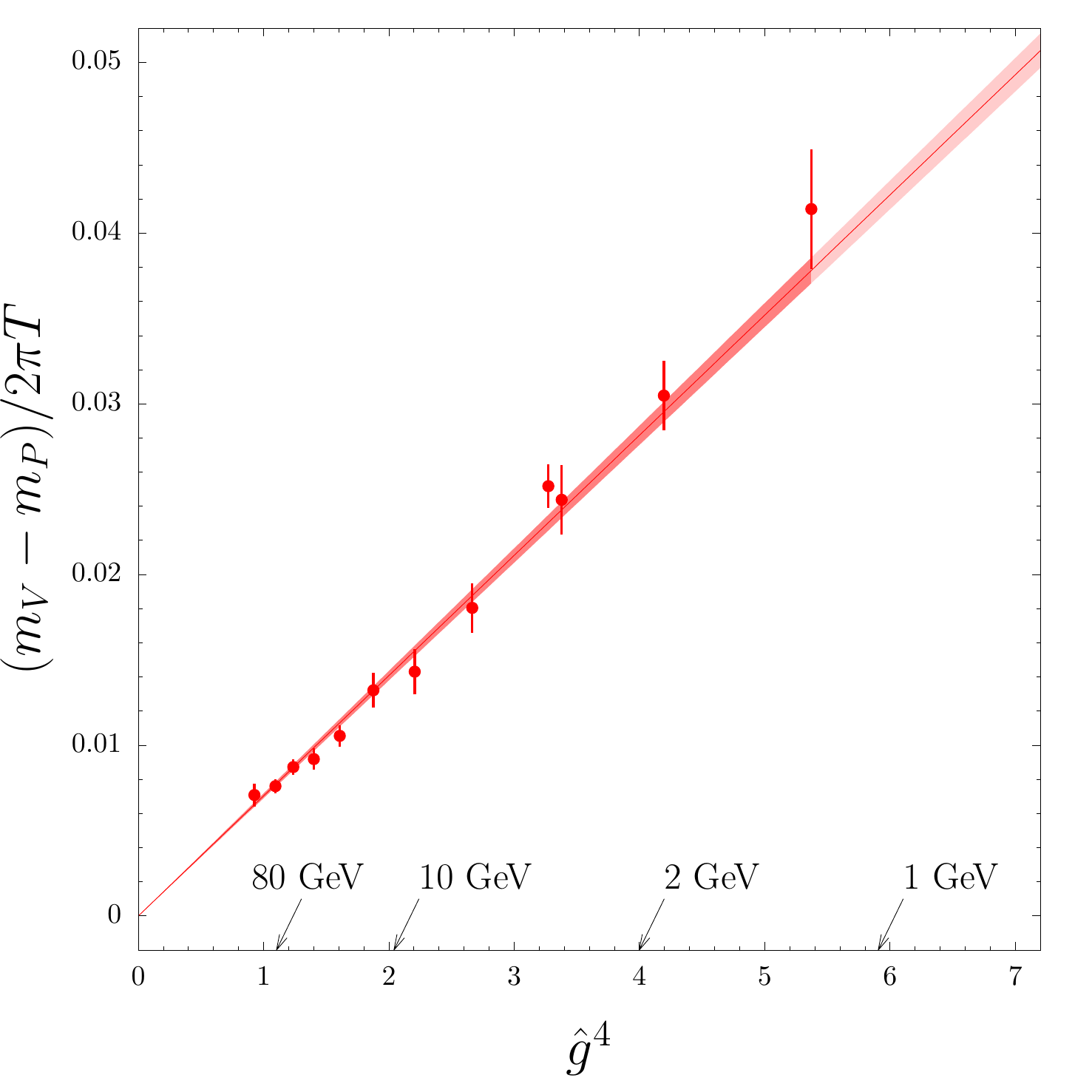}
    \caption{Left: Temperature dependence of the pseudoscalar mass normalized to the free theory value $2\pi T$, after subtracting the known leading terms. Right: the mass difference between the vector and pseudoscalar masses normalized to $2\pi T$. In both cases the temperature dependence is parametrized by plotting as a function of $\hat{g}^4$.}
    \label{fig:spin}
\end{figure}

The pseudoscalar mass has been parametrized with a quartic polynomial in the renormalized coupling
\begin{align}
\label{eq:pion}
    \frac{m_P}{2\pi T} \, = \, p_0 + p_2 \hat{g}^2 + p_3 \hat{g}^3 + p_4 \hat{g}^4 \, .
\end{align}
The leading coefficients $p_0$ and $p_2$ are found to be in agreement with the free theory and the 1-loop order contribution respectively. For $p_3$ and $p_4$ we found $p_3=0.0038(22)$ and $p_4=-0.0161(17)$. The temperature dependence of the pseudoscalar mass is shown in Figure \ref{fig:spin} on the left as a function of $\hat{g}^4$, after subtracting the known leading terms, i.e. the tree-level and the 1-loop perturbative result. The subtracted data have a linear behaviour in $\hat{g}^4$ over more than two orders of magnitude in the temperature.

On the right panel of Figure \ref{fig:spin} we show the temperature dependence of the mass difference between the vector and the pseudoscalar masses, by plotting it as a function of $\hat{g}^4$. By fitting our results with fit ansatz
\begin{align}
    \frac{(m_V-m_P)}{2\pi T} \, = \, s_4 \hat{g}^4
\end{align}
we find $s_4=0.00704(14)$. While the effective field theory predicts these terms to start at $O(\hat{g}^4)$, we have found spin-dependent terms to be of this order over more than two orders of magnitude in the temperature. These terms are still clearly visible even at the highest temperature we simulated, a fact that cannot be explained by the current 1-loop order perturbative calculation in Eq. (\ref{eq:pt}), which predicts the pseudoscalar and the vector masses to be degenerate.

By taking into account the parametrization for the pseudoscalar mass and the one for the spin-dependent terms, the best parametrization for the vector mass is given by the quartic polynomial
\begin{align}
\label{eq:vec}
    \frac{m_V}{2\pi T} \, = \, p_0 + p_2 \hat{g}^2 +p_3 \hat{g}^3 + (p_4+s_4)\hat{g}^4 \, .
\end{align}
\begin{figure}
    \centering
    \includegraphics[width=.5\textwidth]{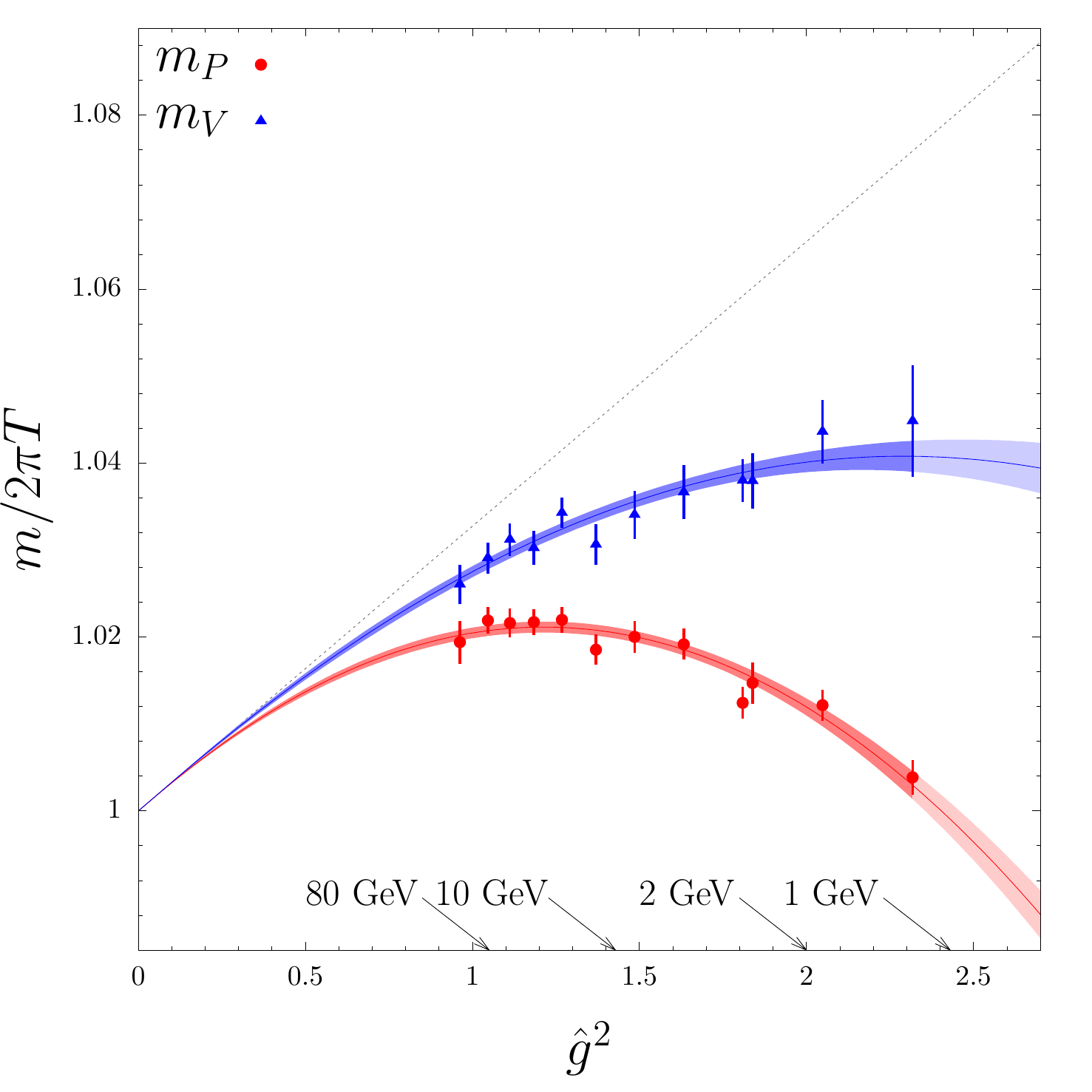}
    \caption{Temperature dependence of the pseudoscalar (red curve) and the vector (blue curve) screening masses as a function of $\hat{g}^2$. The black dashed line represents the 1-loop order perturbative result.}
    \label{fig:temp_dep}
\end{figure}
In Figure \ref{fig:temp_dep} the temperature dependence of the vector and the pseudoscalar masses is shown. We notice that $O(\hat{g}^4)$ terms play a relevant r\^ole in the low temperature regime: On one hand, this contribution explains how the pseudoscalar mass accidentally approaches the free theory value at low temperatures, given the results at $T\sim 1$ GeV. What happens, in fact, at low temperature is that the quartic term in the parametrization of the pseudoscalar mass, see Eq. (\ref{eq:pion}), compensates the $\sim g^2$ term. On the other hand, given the parametrization in Eq. (\ref{eq:vec}), at low temperature the deviation of the vector mass from the free theory result is only due to the spin-dependent term $s^4\hat{g}^4$. These results call for a matching between the effective field theory and QCD at higher order in perturbation theory.
\subsection{Continuum chiral Ward identities and Chiral symmetry restoration}
\label{sec:WI}
At low temperature the axial non-singlet symmetry of the chiral group is spontaneously broken and the axial singlet symmetry is broken by the anomaly. However, when the temperature is high enough, the situation is quite different. On one hand, at high temperature the chiral condensate is expected to drop to zero leading to the restoration of the axial non-singlet symmetry. On the other hand, at very high temperature, since the topological susceptibility is suppressed with the temperature, the topological charge distribution is extremely narrow and peaked at $Q=0$. In terms of the screening spectrum, this restoration pattern translates into the formation of chiral multiplets with degenerate masses. This degeneracy can be made explicit by a set of useful Ward Identities. Assuming there is no spontaneous breaking of chiral symmetry, the following Ward Identities can be easily derived
\begin{align}
\label{eq:WI1}
    \mathcal{O} \, = \, A_\mu^b(z) V_\nu^c(y) \quad &\rightarrow \quad\braket{V_k^a(z) V_k^a(y)} \, = \, \braket{A_k^a(z) A_k^a(y)} \, ,\nonumber\\
    \mathcal{O} \, = \, P^b(z) S^0(y) \quad &\rightarrow \quad2\braket{P^a(z) P^a(y)} \, = \, -\frac{1}{2}\braket{S^0(z) S^0(y)} \, ,\nonumber\\
    \mathcal{O} \, = \, S^b(z) P^0(y) \quad &\rightarrow \quad2\braket{S^a(z) S^a(y)} \, = \, -\frac{1}{2}\braket{P^0(z) P^0(y)} \, ,
\end{align}
where on the left we provide the interpolating operator used to obtain the Ward Indentity and on the right the consequences of such Ward Identity, obtained by taking into accout the variation of the operator $\mathcal{O}$ under axial non-singlet transformations. On the other hand by considering axial singlet transformations, using the same notation, we obtain
\begin{align}
\label{eq:WI2}
    \mathcal{O}\, = \, P^0(z)S^0(y) \quad \rightarrow &\quad \braket{S^0(z) S^0(y)} + \braket{P^0(z) P^0(y)} \, = \, N_f \braket{Q P^0(z) S^0(y)}\, , \nonumber\\
    \mathcal{O}\, = \, P^a(z)S^a(y) \quad \rightarrow &\quad \braket{S^a(z) S^a(y)} + \braket{P^a(z) P^a(y)} \, = \, N_f \braket{Q P^a(z) S^a(y)}\, ,
\end{align}
where the r.h.s vanishes if the topological charge distribution becomes very narrow at large temperature and only the $Q=0$ topological sector contributes to the path integral.

\begin{figure}
    \centering
    \begin{tikzpicture}
            \draw[thin, <->] (0.1,5) -- (3.9,5);
            
            \draw[thin, <->] (0.1,4) -- (3.9,4);
            \draw[thin, <->] (0,3.9) -- (0,0.1);
            \draw[thin, <->] (0.1,0) -- (3.9,0);
            \draw[thin, <->] (4,0.1) -- (4,3.9);
            
            \node[] at (-0.5,4) {$S^a$};
            \node[] at (4.5,4) {$P^a$};
            \node[] at (-0.5,0) {$P^0$};
            \node[] at (4.5,0) {$S^0$};
            
            \node[] at (-0.5,5) {$V^a_\mu$};
            \node[] at (4.5,5) {$A^a_\mu$};

            \node[] at (2,5.4) {\footnotesize{Axial non-singlet}};
            
            \node[] at (-0.4,2) {\rotatebox{90}{{\footnotesize{Axial non-singlet}}}};
            \node[] at (4.4,2) {\rotatebox{-90}{{\footnotesize{Axial non-singlet}}}};

            \node[] at (2,4.4) {\footnotesize{$Q=0$}};
            \node[] at (2,-0.4) {\footnotesize{$Q=0$}};
        \end{tikzpicture}
    \caption{Degeneracy pattern of the mesonic sector of the screening masses if the axial non-singlet symmetry is not spontaneously broken and if only the $Q=0$ topological sector contributes to the path integral.}
    \label{fig:deg_pic}
\end{figure}
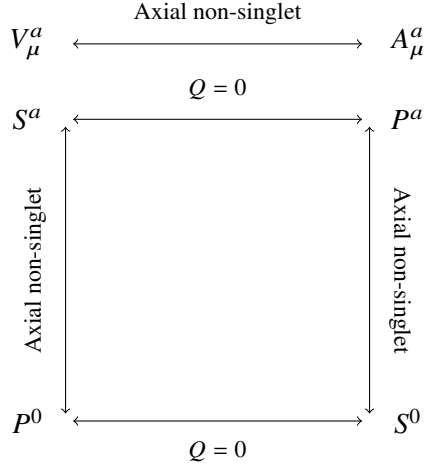

The main consequence of these sets of Ward Identity is the degeneracy of the related screening masses, which produces the standard degeneracy picture represented in Figure \ref{fig:deg_pic}. 

\begin{figure}
    \centering
    \includegraphics[width=0.4\textwidth]{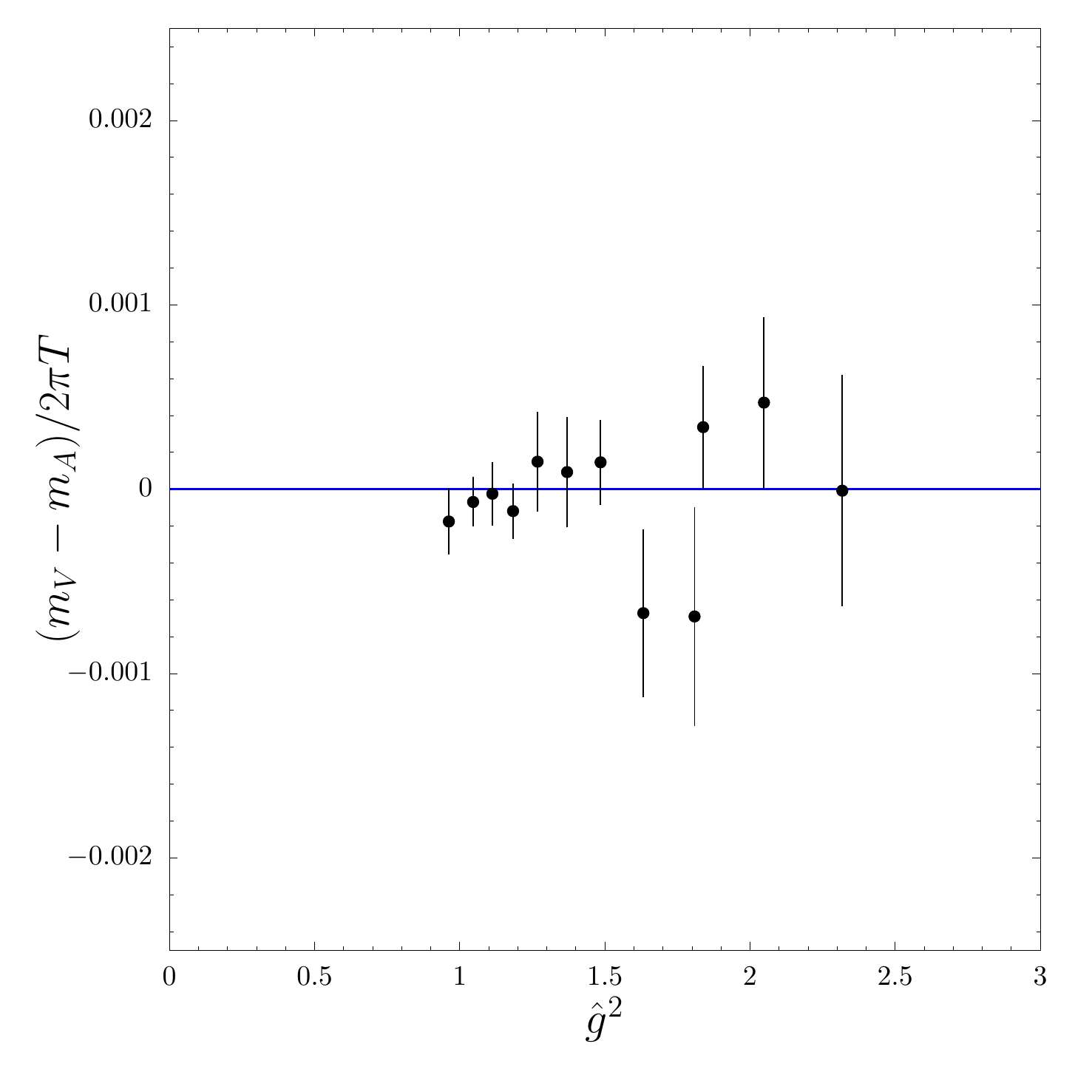}
    \includegraphics[width=0.4\textwidth]{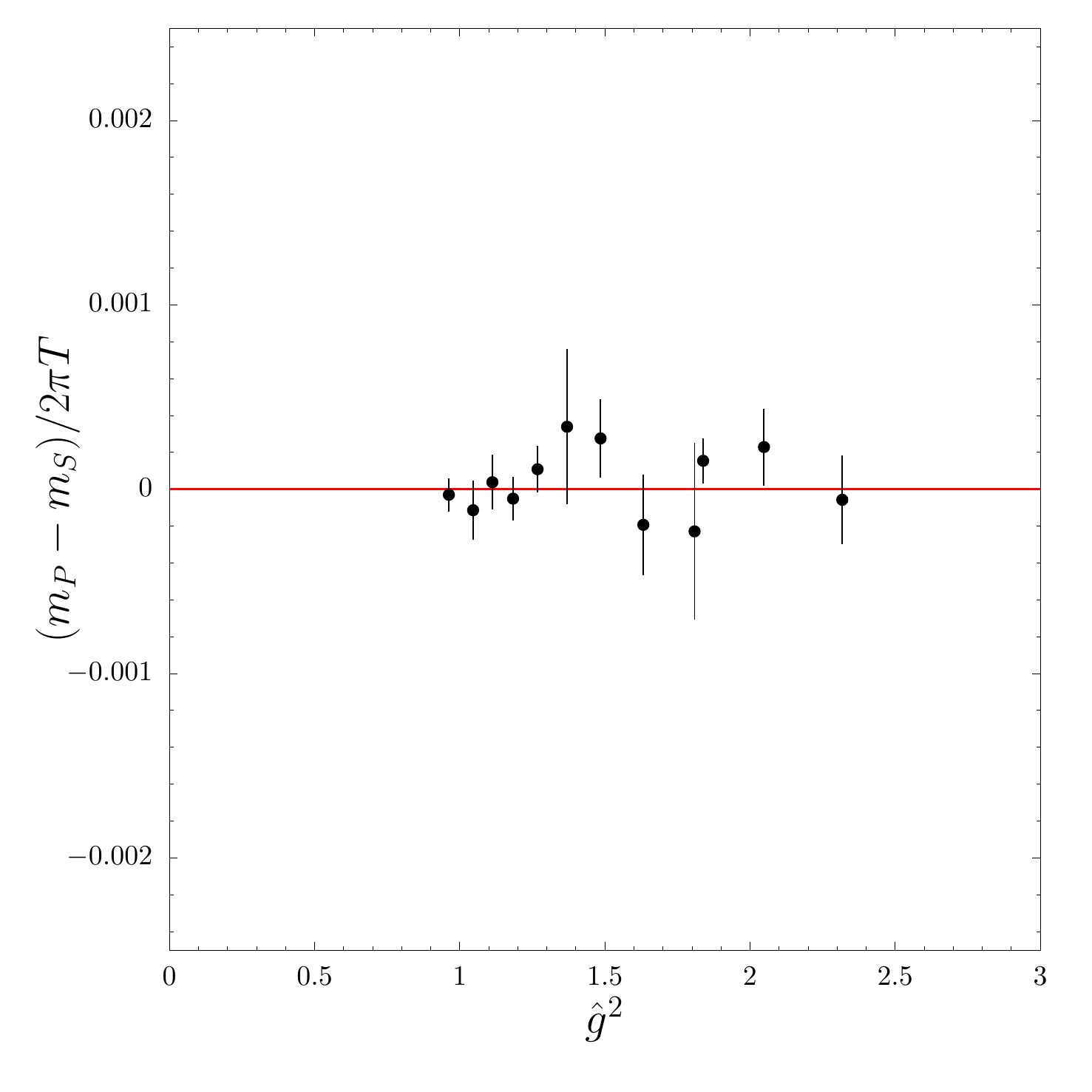}
    \caption{Left: mass difference between the vector and the axial channels. Right: mass difference between the pseudoscalar and the scalar channels.}
    \label{fig:deg}
\end{figure}

For the purpose of this study, the relevant Ward Identities are the first one in Eq. (\ref{eq:WI1}) and the second one in Eq. (\ref{eq:WI2}), i.e. the ones involving only flavor non-singlet interpolating operators.
In Figure \ref{fig:deg} on the left we show the mass difference between the vector and the axial-vector masses as a function of $\hat{g}^2$. For all the temperatures the mass difference is compatible with the expected restoration pattern (blue horizontal line) within the statistical error, which is a clear signal of axial non-singlet symmetry restoration from 1 GeV up to $\sim 160$ GeV. Analogously, on the right we provide the mass difference between the pseudoscalar and the scalar masses. As shown in the figure, the mass difference obtained by restricting the calculation to the zero-topology sector, is compatible with the expected behaviour if $Q=0$ of the r.h.s. of Eq. (\ref{eq:WI2}) (red horizontal line).
\section{Conclusions}
In this talk we have shown how the use of a non-perturbative definition of the renormalized coupling in a finite volume, combined with step-scaling techniques provides a solid strategy to simulate QCD at very high temperature with a moderate computational effort. The applicability of this strategy is also strengthened by the fact that in the high temperature regime finite volume effects are exponentially suppressed for large $LT$.

The successful implementation of this strategy allowed us to study in great detail the mesonic screening spectrum for the first time from $T\sim 1$ GeV up to $\sim 160$ GeV. The temperature dependence of these masses showed a non-trivial behaviour which cannot be explained by the current 1-loop order perturbative result. In particular, $O(\hat{g}^4)$ terms are needed both in the high and in the low temperature regime in order to explain the vector and the pseudoscalar screening masses and their difference. In general, our results are consistent with the effective field theory predictions, however the 1-loop matching is reliable only for temperatures well above the highest temperature we considered.

Moreover, by studying the scalar and the axial-vector channels we observed no signal of chiral symmetry breaking in the entire range of temperature. The numerical results are also supported by Ward Identities obtained in presence of chiral symmetry. The degeneracy between the vector and the axial-vector channel is consistent with the restoration of the axial non-singlet symmetry in the high temperature regime, while the degeneracy between the pseudoscalar and the scalar masses agrees with the fact that only the zero-topological sector contributes to the path integral.
\section*{Acknowledgments}
We acknowledge PRACE for awarding us access to the HPC system MareNostrum4 at the Barcelona Supercomputing Center (Proposals n. 2018194651 and 2021240051) where most of the numerical results presented in this paper have been produced. We also thank CINECA for providing us with computer-time on Marconi (CINECA- INFN, CINECA-Bicocca agreements, ISCRA B projects HP10BF2OQT and HP10B1TWRR). The R\&D has been carried out on the PC clusters Wilson and Knuth at Milano-Bicocca. We thank all these institutions for the technical support.

%\begin{thebibliography}{99}
\bibliographystyle{ieeetr} 
\bibliography{lattice2022}
%\end{thebibliography}

\end{document}